# Plasmonics of Topological Insulators at Optical Frequencies

Jun Yin,[1] Harish N. S. Krishnamoorthy,[2] Giorgio Adamo,[2] Alexander M. Dubrovkin,[2] Yidong D. Chong,[1,2] Nikolay I. Zheludev,[1,2,3] Cesare Soci[1,2,*]

*1. Division of Physics and Applied Physics, School of Physical and Mathematical Sciences, Nanyang Technological University, 21 Nanyang Link, Singapore 637371*

*2. Centre for Disruptive Photonic Technologies, TPI, Nanyang Technological University, 21 Nanyang Link, Singapore 637371*

*3. Optoelectronics Research Centre, University of Southampton, SO17 1BJ, UK*

*Corresponding author: csoci@ntu.edu.sg

**The development of nanoplasmonic devices, such as plasmonic circuits and metamaterial superlenses in the visible to ultraviolet frequency range, is hampered by the lack of low-loss plasmonic media. Recently, strong plasmonic response was reported in a certain class of topological insulators. Here, we present a *first-principles* density functional theory analysis of the dielectric functions of topologically insulating quaternary $(Bi,Sb)_2(Te,Se)_3$ trichalcogenide compounds. Bulk plasmonic properties, dominated by interband transitions, are observed from 2-3 eV and extend to higher frequencies. Moreover, trichalcogenide compounds are better plasmonic media than gold and silver at blue and UV wavelengths. By analysing thin slabs, we also show that these materials exhibit topologically protected surface states, which are capable of supporting propagating plasmon polariton modes over an extremely broad spectral range, from the visible to the mid-infrared and beyond, owing to a combination of inter- and intra-surface band transitions.**

The fields of metamaterials and plasmonics have seen an extraordinary evolution in recent years, from the initial theoretical predictions of artificial negative refractive indices and cloaking[1,2] to the experimental realization of photonic metadevices with various functionalities that can be engineered and obtained on demand.[3,4] However, the practical implementations of plasmonic and metamaterial devices have long been hampered by energy dissipation in plasmonic media, especially in the visible to ultraviolet (UV) range, where even the best plasmonic metals (such as gold, silver and aluminium) suffer from strong dissipation due to interband electronic transitions and Drude losses.[5] This has inspired a search for alternative low-loss plasmonic materials for optical devices operating at high frequencies.[6] Candidates that are currently being investigated include highly doped semiconductors, metallic alloys, nitrides and oxides[7] and, more recently, two-dimensional materials and topological insulator (TI) materials.

Localized plasmons have recently been observed in TIs at THz frequencies (in $Bi_2Se_3$)[8,9] as well as visible-to-UV frequencies (in single crystal $Bi_{1.5}Sb_{0.5}Te_{1.8}Se_{1.2}$ or BSTS)[10]. These experiments observed large bulk resistances and surface-dominated transport.[11,12] In addition, multiple plasmon modes were recently reported in solution-synthesized $Bi_2Te_3$ nanoplates[13] and nano-discs and flakes of BSTS.[14] Ellipsometric data indicated the presence of high optical conductivities, possibly linked to the existence of topologically protected surface states. If true, BSTS or other TIs would be a radically new material platform for broadband nanoplasmonic devices that can be modulated optically[15], through injection of electrons,[16] or by applied magnetic fields[17].

In this work, we use density functional theory (DFT) calculations to investigate the electronic band structures and optical properties of seven compounds in the $Bi_xSb_{1-x}Te_ySe_{1-y}$ (BSTS) family of TIs, in bulk crystal and thin film forms, with the overall goal of determining their suitability for nanoplasmonics (similar DFT studies have proven effective with alkali-noble intermetallics[18] and gallium-doped zinc oxide[19]). Our results show that plasmonic behaviour of quaternary trichalcogenide compounds in the optical part of the spectrum has three origins: a) Bulk interband transitions contributing primarily in the visible spectral region; b) Intraband transition within topologically protected surface bands contributing in the mid-infrared region; c) Interband transition between topologically protected surface states and bulk states, dominating in the UV-NIR range.

**Results and Discussion**

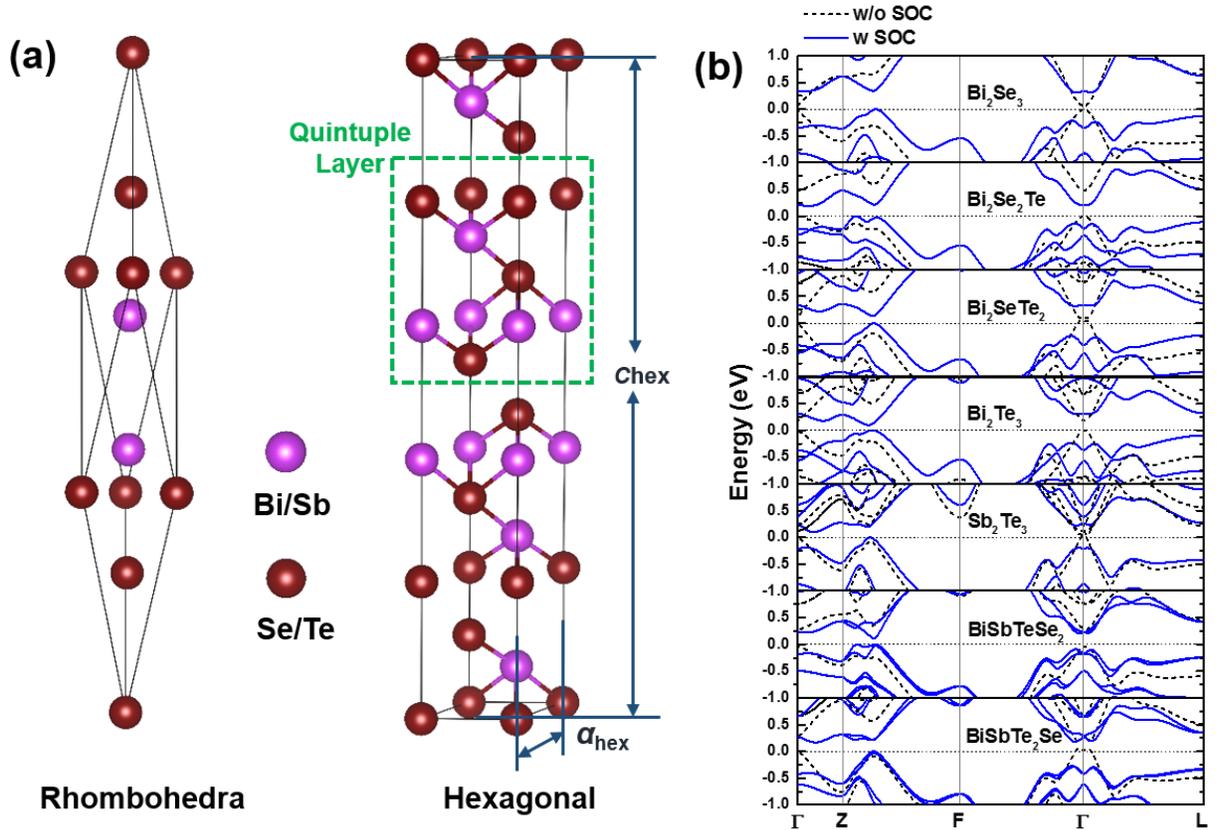

**Figure 1**. (a) Crystal structure of BSTS compounds exhibiting topological insulator behaviour (rhombohedral and equivalent hexagonal lattice cell, with a quintuple layer (QL) shown by the green rectangle); (b) band structures of bulk BSTS compounds calculated with and without account of spin-orbit coupling.

Before studying the optical properties of the BSTS crystals, we optimized their crystallographic structure parameters and calculate their band structures. The three-dimensional bi-chalcogenide compounds have a rhombohedral crystal structure arranged in the order Bi(Sb)-Se(Te)-Bi(Sb)-Se(Te), and all of them belong to space group $R\bar{3}m$.[11] As shown in Fig. 1a, this is a layered structure forming the quintuple layer (QL) unit cell. The intra-layer interaction within a QL can be either covalent or ionic, whereas the weak inter-layer interaction has van der Waals character. We performed DFT calculations of the equilibrium lattice constants within the local-density approximation (LDA), with and without accounting for spin-orbital coupling (SOC). As shown in Supplementary Fig. 2, the calculated lattice parameters are in good agreement with available experimental data[11, 20] Due to the larger ionic radius of Te lattice parameters, $a_{hex}$ and $c_{hex}$ increase with Te content (*e.g.* $Bi_2Te_3$ *vs*. $Bi_2Se_3$). The agreement with experimental data improves when SOC is accounted for, which indicates the validity of our chosen pseudopotentials (see Supplementary Fig. 1).

The calculated band structures of the seven bulk BSTS compounds are shown in Fig. 1b, both with (solid blue lines) and without (black dashes) accounting for SOC. Without SOC, all the compounds exhibit parabolic band dispersions with small (<0.5 eV) direct band gaps at the Γ-point. The direct gaps for the binary compounds $Bi_2Se_3$ (54 meV), $Bi_2Te_3$ (159 meV) and $Sb_2Te_3$ (123 meV), as well as for the ternary compounds $Bi_2Se_2Te$ (477 meV) and $Bi_2Te_2Se$ (84 meV), compare well with previous calculations.[21, 22] The characteristic effect of SOC on the electronic band structures of topological insulators is to produce a band inversion, inducing topologically nontrivial band gaps.[23, 24] Such gaps are indeed observed at the Γ-point for the trichalcogenide compounds (Fig. 1b). The conduction band minima and valence band maxima shift away from the Γ-point, along the symmetric Z-F direction, turning the direct band gaps into indirect band gaps of width 0.10-0.35 eV (see Supplementary Fig. 3). Note that it is possible to estimate the gap energies more accurately by introducing self-energy corrections in the quasiparticle energy approximation, using many-body perturbation theory (e.g., the GW method).[25, 26] However, this would not have noticeable effects on the visible to near-infrared optical properties, which are dominated by high-energy interband transitions.

To quantify the plasmonic properties of BSTS compounds, we separately evaluated their bulk and surface dielectric functions. In the bulk, the complex dielectric function (ε) originates from interband transitions, which we calculated from the band structures using the Bethe-Salpeter and Kramers-Kronig equations (see *Computational Methods*). In Fig. 2, we plot the real part of the dielectric function (ε'), which describes the strength of the polarization induced by an external electric field, and the imaginary part (ε"), which describes the losses encountered in polarizing the material.[5] The strongest peak in the spectral dispersion of ε" moves from the near-infrared (0.5-1.5 eV) to the visible spectral region (1.5-2.5 eV) when tellurium content is reduced. This peak arises from interband transitions occurring along the symmetry line near the Γ-point (0, 0.16666, -0.16666). Once SOC effects are included in the optical response calculations, we find that there are two main absorption peaks in ε" (see Supplementary Fig. 5). The first peak in the visible region (1.5-2.0 eV) is red-shifted as the tellurium content is reduced, which is attributable to the resonant character of interband transitions occurring along the symmetry line near the Γ-point. The other strong peaks appear at ~0.6 eV due to interband transitions around the (0.33333, -0.5, -0.5) point. The transition details for the optical permittivity are listed in Supplementary Table 1. The calculated values of bulk optical permittivity agree well with experimental data reporting prominent peaks of the imaginary part of the permittivity between 1.5 and 2 eV for $Bi_2Te_3$, $Sb_2Te_3$ and $Bi_{1.5}Sb_{0.5}Te_{1.8}Se_{1.2}$ (see Fig. 2b),[10, 13, 27] although calculated permittivities at peaks are somewhat higher than the experimentally measured ones. For the real part of the optical permittivity, the main features are broad peaks centred around 1.0-2.0 eV, followed by a steep decrease between 1.5 and 3.0 eV, after which ε' becomes negative and eventually increases slowly toward zero at higher energies. The epsilon-near-zero behaviour, interesting for many applications[28, 29, 30] appears in the UV region, above 5 eV.

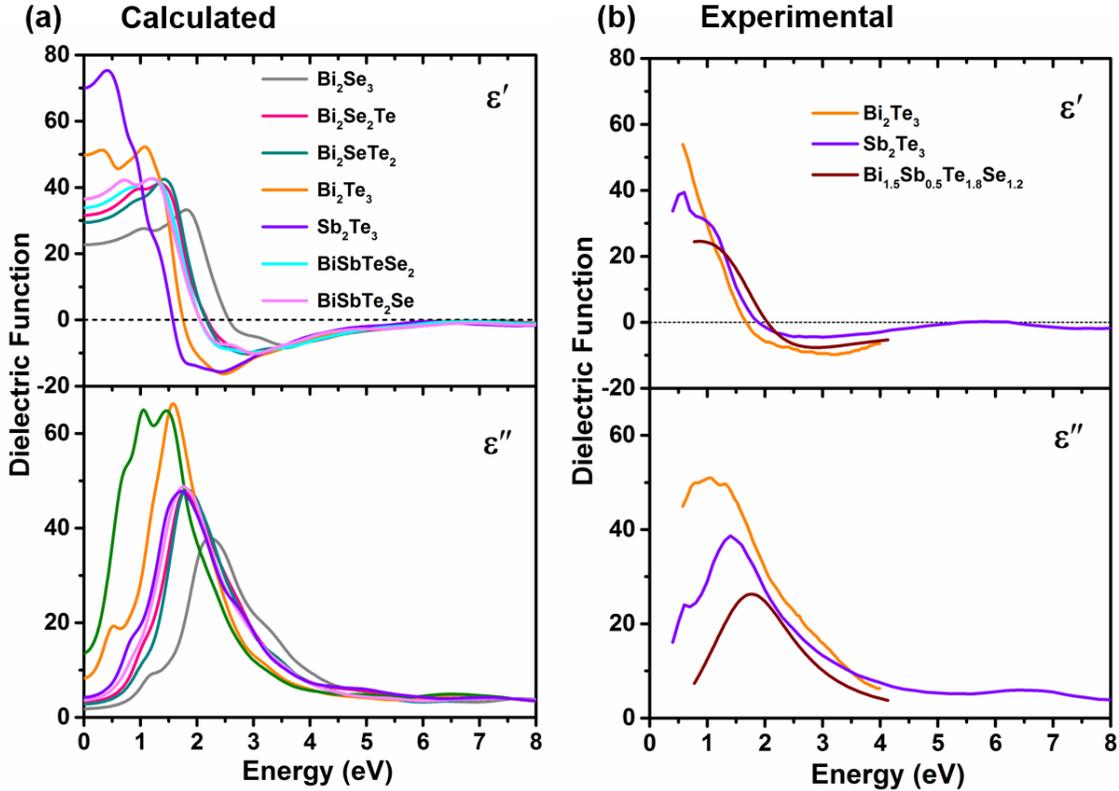

**Figure 2.** Interband transition contribution to the real (ε') and imaginary part (ε") of the permittivity of bulk BSTS materials: (a) DFT results without SOC and (b) experimental data obtained from refs. [10, 13, 27].

Thus, for the bulk materials, plasmonic behaviour emerges when ε' becomes negative in a frequency range of around 2 eV or higher, depending on the compound. We note that negative permittivities due to interband transitions have also been reported in previous DFT calculations on $Bi_2Se_3$.[31] Our results show that increasing the tellurium content gives plasmonic behaviour over a wider frequency range, with much larger negative values of ε' (e.g. for $Bi_2Te_3$).

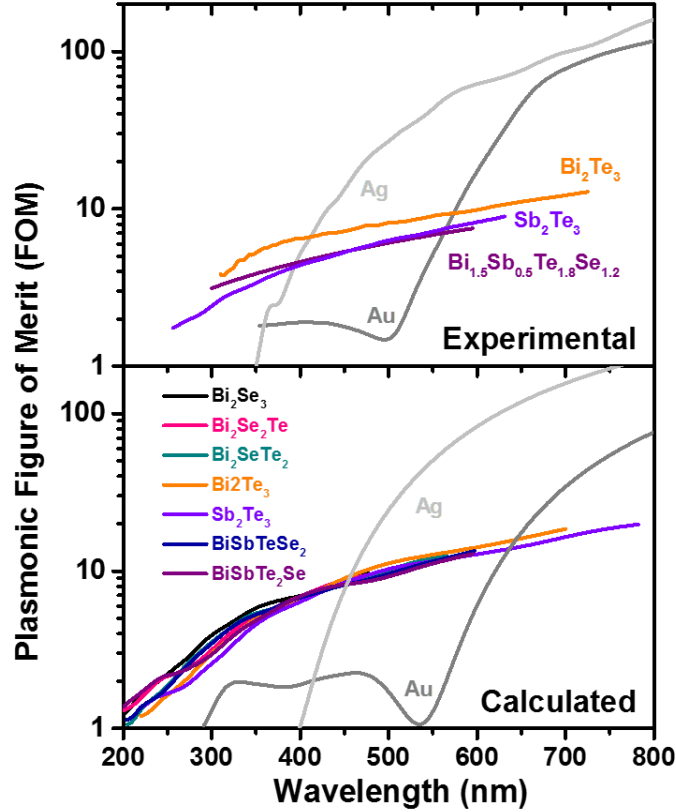

**Figure 3.** Plasmonic figures of merit for bulk BSTS compounds and noble metals determined using experimental dielectric constants from the literature (top panel) and theoretical dielectric constants obtained from DFT calculations (bottom panel).

The bulk dielectric functions indicate that the BSTS compounds can outperform plasmonic materials in the visible to UV spectral range, including traditional plasmonic metals. To quantify this, in Fig. 3, we determine the plasmonic figure of merit,[32] $FOM = \frac{L_{SPP}}{\lambda_{SPP}} = \frac{\text{Re}(k_{SPP})}{2\pi \text{Im}(k_{SPP})}$, where $k_{SPP} = k_0 \sqrt{\frac{\varepsilon_{TI}\varepsilon_{air}}{\varepsilon_{TI}+\varepsilon_{air}}}$ is the complex wave vector of the surface plasmon polariton (SPP) mode. Such FOM can be thought of as the number of wavelengths an SPP mode can travel before dissipation. In Fig. 3 we plot the FOM of bulk BSTS compounds obtained from our DFT calculations as well as experimental data available in the literature.[10, 13, 27] FOM of silver and gold are also included for comparison. Both sets of data indicate that the BSTS materials have higher FOM than gold below ~600 nm, and higher FOM than silver below ~450 nm. Most strikingly, in the UV-blue wavelengths of ~400 nm, SPP modes in bulk BSTS chalcogenide crystals are expected to propagate 3 to 6 wavelengths more than in Au or Ag.

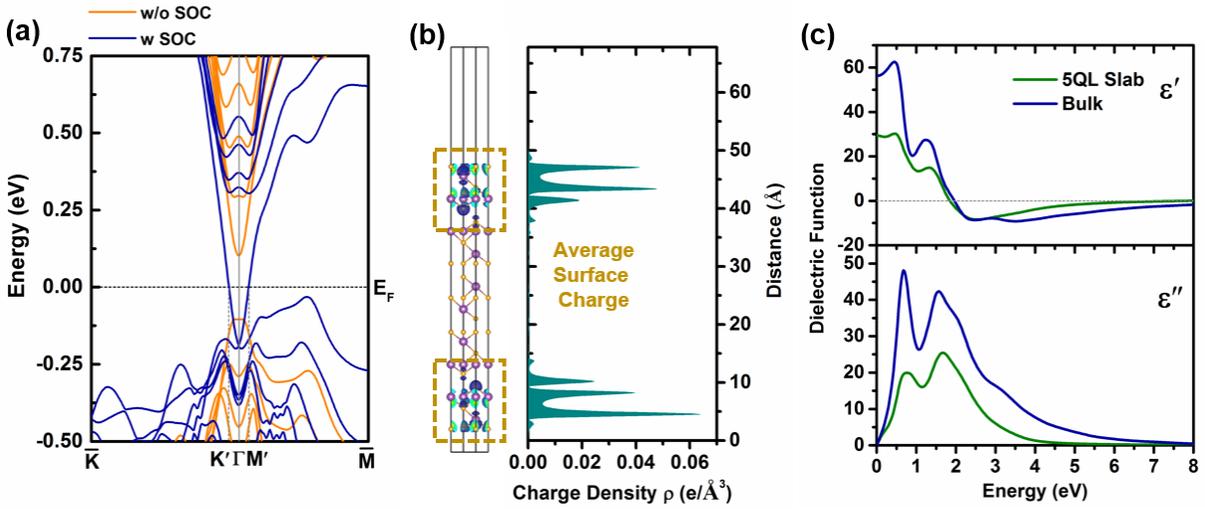

**Figure 4.** (a) Band structure of 5QL $Bi_2Se_3$ slab without and with inclusion of spin-orbit coupling effects (the horizontal dashed line indicates the Fermi energy level and K′-M′ the region of occupied surface states). (b) Three- and one-dimensional charge density distribution of conducting surface states; all occupied orbitals below the Fermi level and their degeneracy (see Supplementary Fig. 7) were considered to calculate the average charge density. (c) Comparison between real and imaginary parts of the dielectric functions of bulk and 5QL $Bi_2Se_3$ slab with SOC.

We now turn our attention to the surface properties of BSTS compounds. To study the optical response of topological surface states, we performed DFT calculations considering thin slabs. The emergence of topological effects and charge accumulation at surface states in TIs is expected to be clearly seen in slabs with thickness of a few quintuple layers, as indicated by previous theoretical studies[24, 33] and experimental observations.[34] Based on this, we carried out DFT calculations for a (111) crystal slab with five quintuple layers (5QL) of approximate thickness of ~5 nm, and looked for the emergence of surface states upon activation of spin-orbital coupling. The contribution of surface states to the optical response was then obtained considering two types of transitions: *i. intraband* transitions described by 2D Drude-like response of surface free carriers,[8, 9, 35] and *ii. interband* transitions from topological surface states to higher energy bands.

**Table 1.** Surface state thickness *s*, effective mass *m*, two-dimensional carrier concentration $n_{2D}$ of surface states, carrier scattering time τ, damping factor Γ, and conductivity σ of 5QL-TI slabs.

| Surface | | Effective Mass $m\ (\times m_0)$ | Carrier Concentration $n_{2D}(cm^{-2})$ | Fermi Energy $E_f$ (eV) | Surface Layer Thickness $d$ (nm) | Carrier Scattering Time τ (fs) | Damping Factor $\Gamma = 1/\tau$ (eV) | DC conductivity $\sigma_{2D}^{DC}$ (S/m) |
|---|---|---|---|---|---|---|---|---|
| Bi$_2$Se$_3$ | DFT | 0.156 | 6.33×10$^{14}$ | 0.189 | 0.92 | - | - | 1.42×10$^6$ |
| | Exp | 0.13-0.15 (refs. 36, 37) | 3.0±0.2×10$^{13}$ (ref. 38) | - | - | 55-150 (refs. 36, 39) | 0.028-0.075 | 4.54×10$^4$ (ref. 40) |
| Bi$_2$Se$_2$Te | DFT | 0.089 | 2.97×10$^{13}$ | 0.171 | 1.27 | - | - | 9.32×10$^5$ |
| Bi$_2$SeTe$_2$ | DFT | 0.126 | 7.91×10$^{13}$ | 0.259 | 1.28 | - | - | 1.40×10$^6$ |
| | Exp | 0.11 (ref. 41) | 1.50×10$^{12}$ (ref. 39) | - | - | 48 (ref. 41) | 0.086 | 1.00×10$^5$ (ref. 42) |
| Bi$_2$Te$_3$ | DFT | 0.178 | 8.74×10$^{13}$ | 0.194 | 1.90 | - | - | 7.07×10$^5$ |
| | Exp | - | 7.00×10$^{12}$ (ref. 43) | - | - | 54 (ref. 44) | 0.077 | 1.65×10$^5$ (ref. 45) |
| Sb$_2$Te$_3$ | DFT | 0.095 | 5.67×10$^{13}$ | 0.103 | 1.79 | - | - | 3.99×10$^5$ |
| BiSbTeSe$_2$ | DFT | 0.087 | 7.74×10$^{14}$ | 0.109 | 1.20 | - | - | 6.29×10$^5$ |
| BiSbTe$_2$Se | DFT | 0.110 | 9.56×10$^{14}$ | 0.216 | 2.05 | - | - | 7.30×10$^5$ |
| | Exp | 0.32 (ref. 46) | 3.80×10$^{13}$ (ref. 35) | - | - | 58 (ref. 46) | 0.071 | 4.57×10$^4$ (ref. 47) |

Here we take Bi$_2$Se$_3$ slabs as an example and present their electronic structure, charge distribution of the surface state, and optical permittivity (Fig. 4). Similar data for the other BSTS compounds are provided in Supplementary Figs 6-9. Without accounting for SOC effects, the band structure of 5QL-slab (orange solid line in Fig. 4a) is similar to those of the bulk crystals obtained using periodic boundary conditions (Fig. 1b). Increasing the slab thickness decreases the energy band gap because of the reduction in quantum confinement, with the energy gap approaching the bulk value at about 19QL thickness (Supplementary Fig. 4). Upon including SOC effects, a single non-degenerate Dirac point appears at the Γ-point. At the Dirac cone below the Fermi energy level, the electronic charge density distribution is mainly localized at the surface and penetrates ~1-2 nm into the bulk (see surface layer thickness in Table 1 and integrated charge density function in Supplementary Fig. 8). The optical response of the 5QL $Bi_2Se_3$ slab shows similar spectral features as the bulk: the dominant peak of the imaginary part of the permittivity appears at 1.6 eV, and a low-energy peak emerges around 0.6 eV in agreement with the experimental dispersion curves shown in Fig. 2b.

Knowledge of the band structure and surface charge distribution of the TI slabs allows extraction of other relevant parameters of the topological insulator, such as effective mass $m$ of the carriers, surface layer thickness $d$ and the surface carrier density $n_{2D}$ used to determine the DC conductivity, $\sigma_{2D}^{DC}$. These parameters are listed in Table 1. For the 5QL-slabs, the effective mass of electron was estimated by fitting the linear dispersion curve of the lowest conduction band with the relationship $m = \frac{\hbar k}{v_F}$, where $v_F$ is the Fermi velocity,[48] while the total surface carrier concentration was obtained by integrating the occupied surface states charge below Fermi energy level (the region K′-M′, see Supplementary Figs 7-8). The intraband surface contribution to the dielectric response was determined using the methodology developed for other two-dimensional systems, such as graphene.[49] The DC conductivity due to the topological insulator surface states is given by $\sigma_{2D}^{DC} = \frac{1}{d}\frac{e^2 E_f \tau}{\pi \hbar^2}$, where $E_f$ is the Fermi energy, $\tau$ is the carrier scattering time and $d$ is the thickness of the surface layer. At higher frequencies, the dispersion of the optical conductivity is given by the two-dimensional (2D) Drude model:[9, 35, 49, 50] $\sigma_{2D}^{AC}(\omega) = \frac{\sigma_{2D}^{DC}}{(1-i\omega\tau)}$. The contribution of surface free carriers to the optical response can then be evaluated from the expression: $\varepsilon_{2D}(\omega) = \frac{i\sigma_{2D}^{AC}(\omega)}{\varepsilon_0 \omega} = \frac{i}{\varepsilon_0 \omega}\frac{\sigma_{2D}^{DC}}{(1-i\omega\tau)}$. This expression coincides with the dispersion of the dielectric constant of conventional plasmonic metals, provided that the three dimensional (3D) static conductivity, $\sigma_{3D}^{DC}$, is replaced by the 2D static conductivity, $\sigma_{2D}^{DC}$, of the surface states. Unlike the case of conventional plasmonic metals, however, this dispersion relationship has no cut-off (plasma) frequency, thus yielding negative real permittivity even in the optical spectral range, irrespectively of the carrier density (Fermi energy). The contribution of surface state intraband transitions to the dielectric constant (Figs. 5b-h, dotted lines) was quantified using the parameters determined from first principles DFT calculations, except for the free carrier damping factor, assumed to be Γ=0.07 based on the typical experimental carrier lifetime of ~60 fs (see Table 1)[36, 41, 44, 46, 51]. The plasmonic contribution of surface free carriers is rather small in the visible part of the spectrum but becomes prominent at longer wavelengths, starting from the near IR region. This is reflected in the extremely high plasmonic figures of merit of BSTS TI slabs at the longer wavelengths (refer to Supplementary Fig. 10a).

The dielectric response arising from interband surface transitions, that is optical transitions from topologically protected surface state to higher energy bands, is shown if Fig. 5 (solid lines). The real and imaginary part of the dielectric constants were evaluated from the DFT band structure of the TI slabs using similar methodology employed for estimating the optical response of the bulk (as described in *Computational Methods*). These transitions give rise to plasmonic response (negative real permittivity) in the optical part of the spectrum with strong absorption features around 1.5 eV and 0.5 eV. In contrast to the bulk, these surface interband transitions yield plasmonic behaviour over an extremely broad spectral range, from the visible through the NIR, with lower losses and appreciably high plasmonic figures of merit (see Supplementary Fig. 10b).

Our estimate of the overall optical response arising from intra- and inter-band transitions from topological surface states of BSTS compounds is in good qualitative agreement with the optical permittivity of the surface of $Bi_{1.5}Sb_{0.5}Te_{1.8}Se_{1.2}$ determined from ellipsometric measurements (Fig. 5a), assuming a multilayer model with a thin Drude surface layer.[10] This confirms that topologically protected surface charge carriers have appreciable contribution to the plasmonic response of the TI crystals at optical frequencies, with strong Drude response in the infrared region and significant interband surface contribution throughout the optical spectrum, up to the UV. The good agreement with experiments gives us confidence that the proposed TI slab model, in which the contributions of the conducting surface layer and the bulk are treated independently, can adequately describe the optical and plasmonic response of this family of TI materials. Most importantly, the cumulative plasmonic response of the topological insulator brought about by the combination of inter- and intra-band surface transitions is extremely broadband in nature. Both types of transitions originate from topological surface states that, besides being immune to scattering from defects, are spin-polarised and strongly coupled to the bulk. This opens up exciting opportunities for broadband plasmono-spintronic devices that can be controlled by charge injection, external magnetic fields, or light helicity.

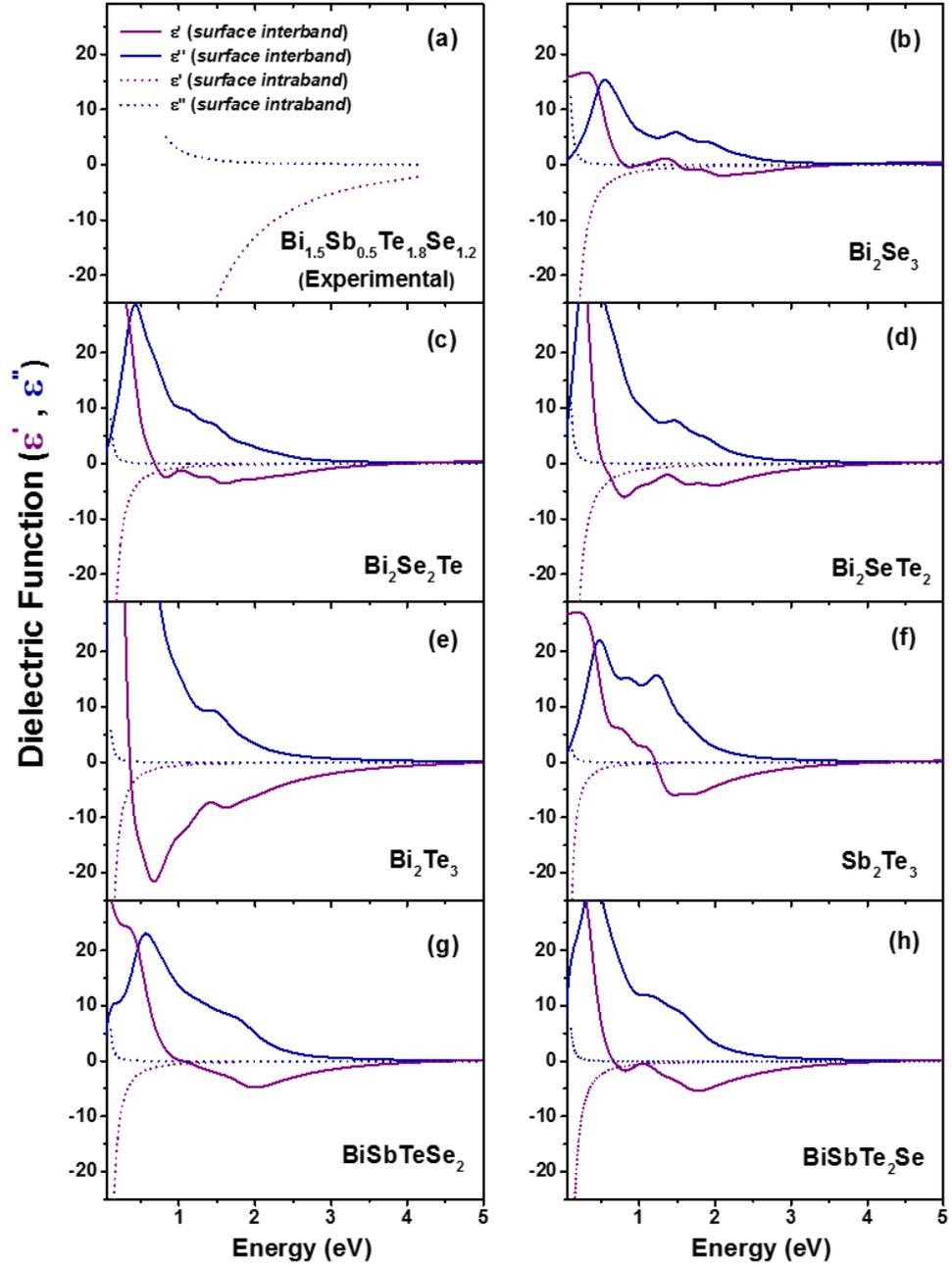

**Figure 5.** Inter- and intra-band contributions to the real and imaginary parts of the optical constants of BSTS TI slabs originating from topological surface states. (a) Experimental values of $Bi_{1.5}Sb_{0.5}Te_{1.8}Se_{1.2}$ surface contribution were derived in ref. [10] using a two-layer model. (b-h) Intraband contributions calculated according to a 2D Drude model (dotted lines) and interband contributions due to transitions from the surface state band to higher energy bands (solid lines).

In summary, we presented a systematic study of the optical and plasmonic properties of BSTS topological insulator crystals using first principles DFT calculations. In addition to their negative bulk permittivity, topological insulators show robust, spin-polarized surface states, which are capable of supporting surface plasmons over a very broad wavelength range. Comparison of the calculated dielectric functions of bulk crystals revealed the dependence of the optical band gap and permittivity on composition, indicating plasmonic behaviour with figures of merit higher than noble metals in the UV-blue spectral region. Increase of Te versus Se content can extend the overall plasmonic response deeper into the NIR region and reduce losses. Furthermore, knowledge of surface charge distribution in thin TI slabs has allowed isolating the plasmonic response of the surface. The combination of *interband* and *intraband* (2D Drude) transitions involving topologically protected surface states yields plasmonic response over an extremely broad spectral range. This new understanding of bulk and surface state contributions to the overall plasmonic response of topological insulators will enable targeted strategies to exploit these materials in nanophotonic applications, for instance tailoring the growth of thin films by epitaxial growth methods to maximise the contribution of topological surface states and lower plasmonic losses at desired optical frequencies.

**Computational Methods**

Density functional theory (DFT) calculations were used to study the electronic structure and optical response of TI materials employing the local density approximation (LDA), using the Quantum ESPRESSO code.[52] Experimental lattice parameters of bulk TI materials[11, 20] were used as the initial structure, and ground states geometries of the TI systems were obtained by the total energy minimization method upon relaxing their crystal framework and atomic coordinates. Electron-ion interactions (plane waves and core electrons) were generated by norm-conserving non-relativistic and relativistic pseudopotentials with electrons for Bi (6s2, 6p3, 5d10); Sb (5s2, 5p3); Se (4s2, 4p2); Te (5s2, 5p2, 4d10). The relativistic effects and spin-orbital coupling (SOC) have significant effects on the band structure due to heavy elements, such as Bi, Sb, and Te. Supplementary Fig. 1 shows the small energy differences (<0.005 Ry) between all-electron and our generated pseudopotentials, confirming the good transferability properties of these pseudopotentials. Single-particle wave functions (charges) were expanded on a plane-wave basis set up to a kinetic energy cutoff of 80 Ry (500 Ry) for both TI bulk and slabs. The bulk and thin film layers were relaxed until forces on the atoms were lower than 0.01 eV/Å. The k-space grid of 6×6×6 in the Brillouin zone was chosen for the calculation of the band structures.

The optical response calculations were performed by the Bethe-Salpeter equations (BSE) method with the YAMBO code, using ground state wavefunctions from Quantum ESPRESSO:[53, 54]

$$(E_{ck} - E_{vk})A^S_{vck} + \sum_{k'v'c'} \langle vck|K_{eh}|v'c'k'\rangle A^S_{v'c'k'} = \Omega^S A^S_{vck}$$

where $E_{ck}$ and $E_{vk}$ are the quasiparticle energies of the conduction and valence states, respectively; $A^S_{vck}$ are the expansion coefficients of the excitons, and $\Omega^S$ are the eigenenergies.

The imaginary part of the permittivity was calculated by evaluating direct electronic transitions between occupied and higher-energy unoccupied electronic states as obtained from

$$\varepsilon'(\omega) \propto \sum_S \left| \sum_{cvk} A^S_{vck} \frac{\langle ck|p_i|vk\rangle}{\epsilon_{ck} - \epsilon_{vk}} \right| \delta(\Omega^S - \hbar\omega - \Gamma)$$

where $\langle ck|p_i|vk\rangle$ are the dipole matrix elements for electronic transitions from valence to conduction states. The real part can then be calculated via the Kramers-Kronig relation $\varepsilon'(\omega) = 1 + \frac{2}{\pi} P \int_0^\infty \frac{\varepsilon''(\omega')\omega' d\omega'}{\omega'^2 - \omega^2}$. SOC interactions and the spinor wave functions were included as input for the electronic and optical calculations on the level of many-body perturbation theory. Previously this approach has been used to successfully study the optical properties of systems with strong spin-orbital interaction.[55] For optical permittivity calculations, a k-point grid corresponding to 12×12×12 and 40 conduction bands and 40 valence bands both for bulk and slabs were chosen to ensure a sufficient description of the optical permittivity.

For the crystal slabs, the optical permittivity consists of interband contribution (including transitions between two surface bands and bulk bands, as well as transitions from surface valence band to bulk conduction bands) and Drude-like free-electron intraband contribution from the metallic surface, which can be described by a two-dimensional Drude model determined by the Fermi energy, carrier scattering time and thickness of the surface as described in the main text.

**Acknowledgments**

Authors acknowledge financial support of the Singapore Ministry of Education (grants MOE2011-T3-1-005 and MOE2013-T2-044) and EPSRC (U.K.) grant EP/M009122/1.

**Author contributions**

C.S. and Y.J. generated the idea and designed the simulation work. Y.J. performed all DFT modelling. H.N.S.K. analyzed the plasmonic response. All the authors contributed to interpretation of the results and writing of the manuscript. C.S. and N.I.Z. supervised the work.

**Additional information**

Supplementary Information accompanies this paper.
The authors declare no competing financial interests.

# Supplementary Information

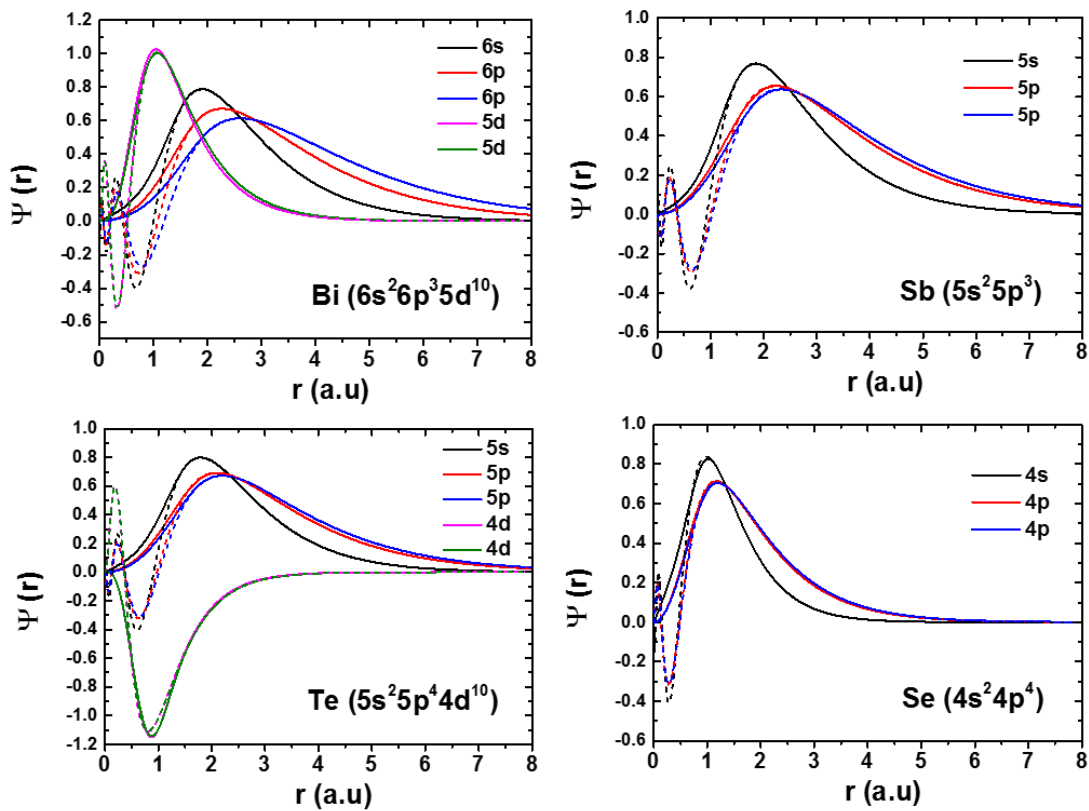

**Supplementary Figure 1.** Comparison of the all-electron wavefunctions with pseodowavefunction of Bi, Sb, Te, and Se atoms.

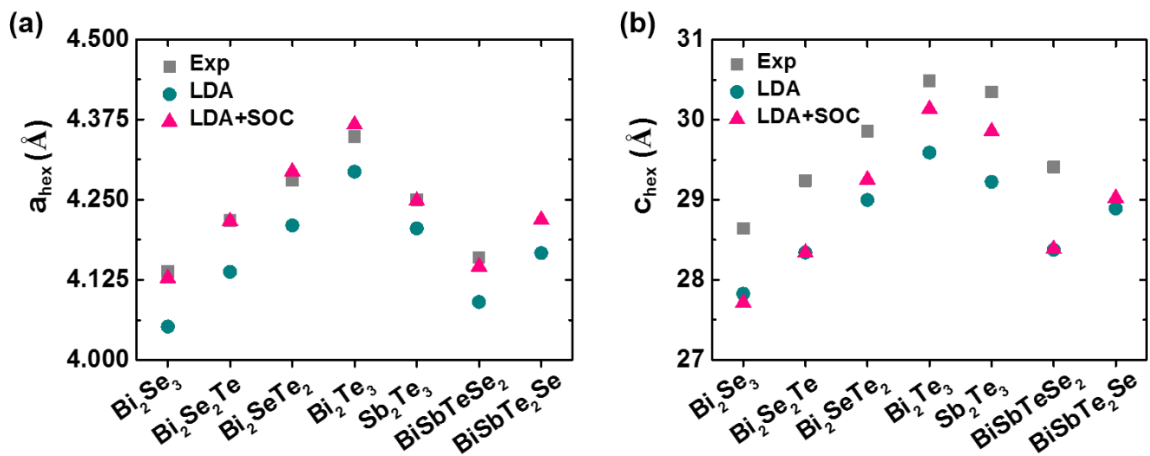

**Supplementary Figure 2.** Lattice parameters $a_{hex}$ (a) and $c_{hex}$ (b) of the hexagonal lattice cell calculated by LDA and LDA+SOC method.

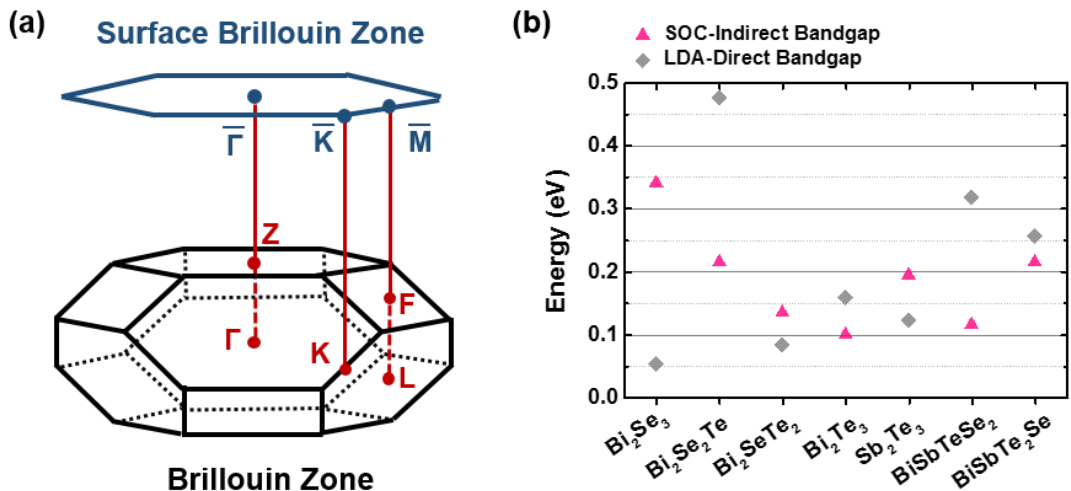

**Supplementary Figure 3.** (a) 3D Brillouin Zone of the bulk TI compounds and 2D projection of the (111) surface; (b) Direct and indirect electronic band gap energies determined from the band structures.

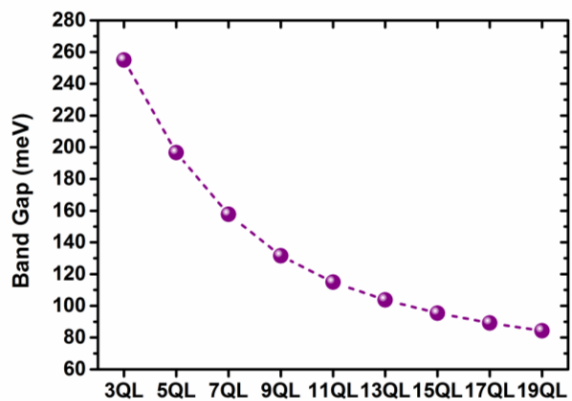

**Supplementary Figure 4.** Bandgap energies of Bi$_2$Se$_3$ slabs with 3 to 19 quintuple layers (3QL to 19QL).

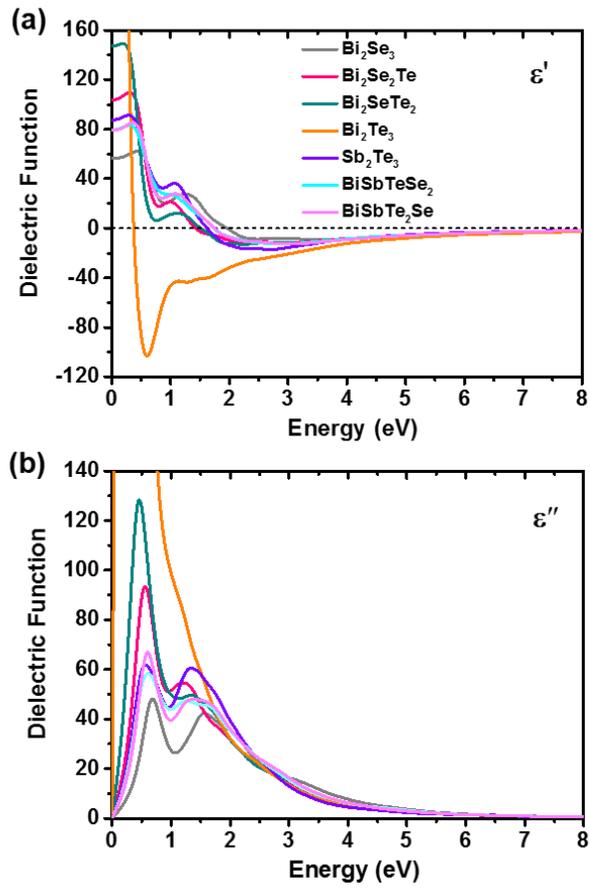

**Supplementary Figure 5.** Bulk interband transition contribution to the real (ε') and imaginary part (ε") of the permittivity of BSTS topological insulator crystals calculated by the BSE method with inclusion of spin-orbital coupling (SOC) effects.

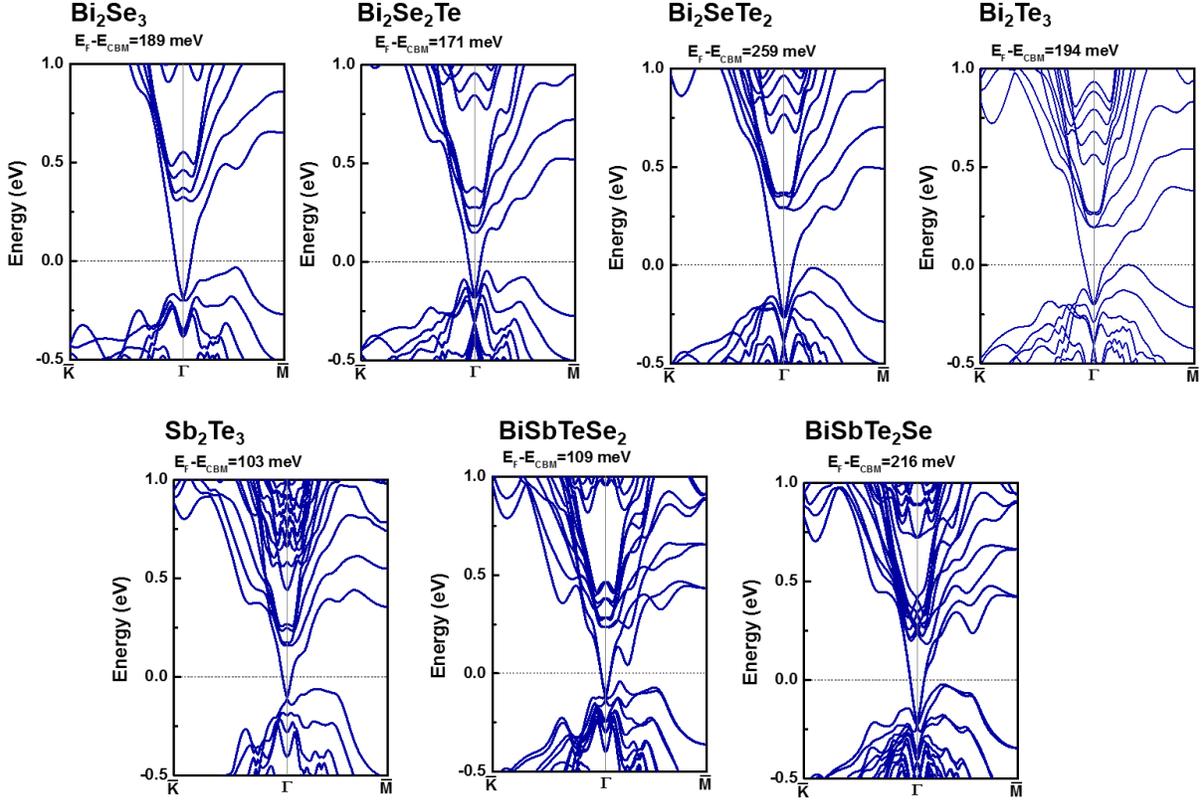

**Supplementary Figure 6.** Band structures of 5QL-slabs of BSTS crystals including SOC effects. Energy scales are centered at Fermi levels (dotted lines), and Fermi energy values relative to the conduction band minimum, $E_\text{F}$-$E_\text{CBM}$, are indicated on top of each panel.

**Supplementary Table 1**. Dominant transitions contributing to the peaks of the imaginary part of the optical permittivity ($\varepsilon''$) of bulk $Bi_2Se_3$.

| Peak | Energy (eV) | Dominant Transitions | Direct Transition K-point |
|---|---|---|---|
| LDA | 2.051 | VBM->CBM | (0, 0.16666, -0.16666) |
| LDA+SOC | 1.579<br>0.677 | VBM-3->CBM, VBM-2->CBM+1<br>VBM-1->CBM, VBM->CBM+1 | (0.33333, -0.5, -0.5) |

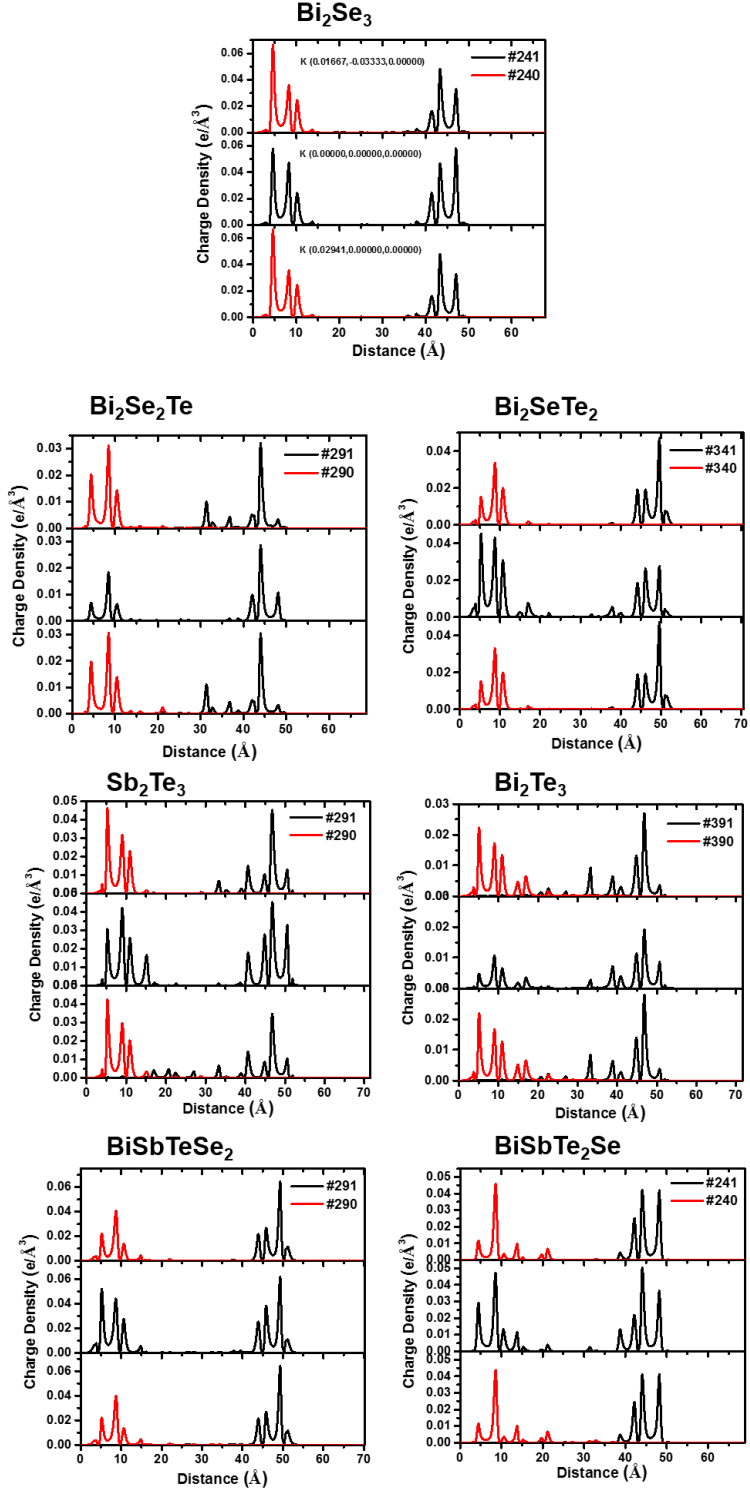

**Supplementary Figure 7.** The uniform charge density distribution of conducting surface state below Fermi energy level at K-points (0.017,-0.333,0.000), Γ(0.000,0.000,0.000) and (0.029,0.000,0.000) for BSTS crystal slabs (black and red lines indicate the degenerated orbitals and numbers indicate the orbital indexes).

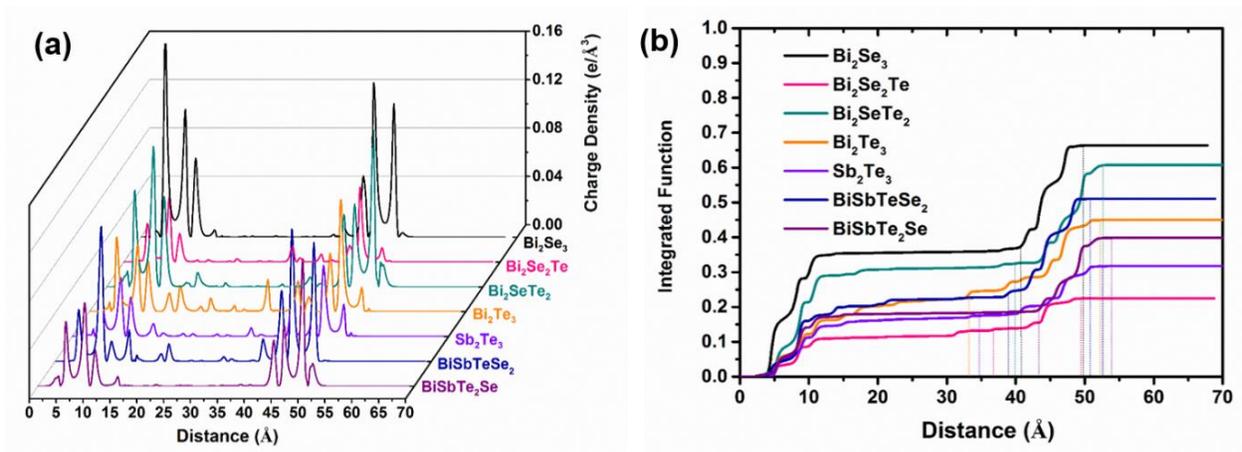

**Supplementary Figure 8.** (a) Total charge density distribution of conducting surface states (in the region Κ′-Γ-Μ′) in thin slabs of the different BSTS compounds; and (b) corresponding integrated charge density function. The top surface state thicknesses for BSTS compounds are determined from vertical dashed lines with 95% increment of total charges.

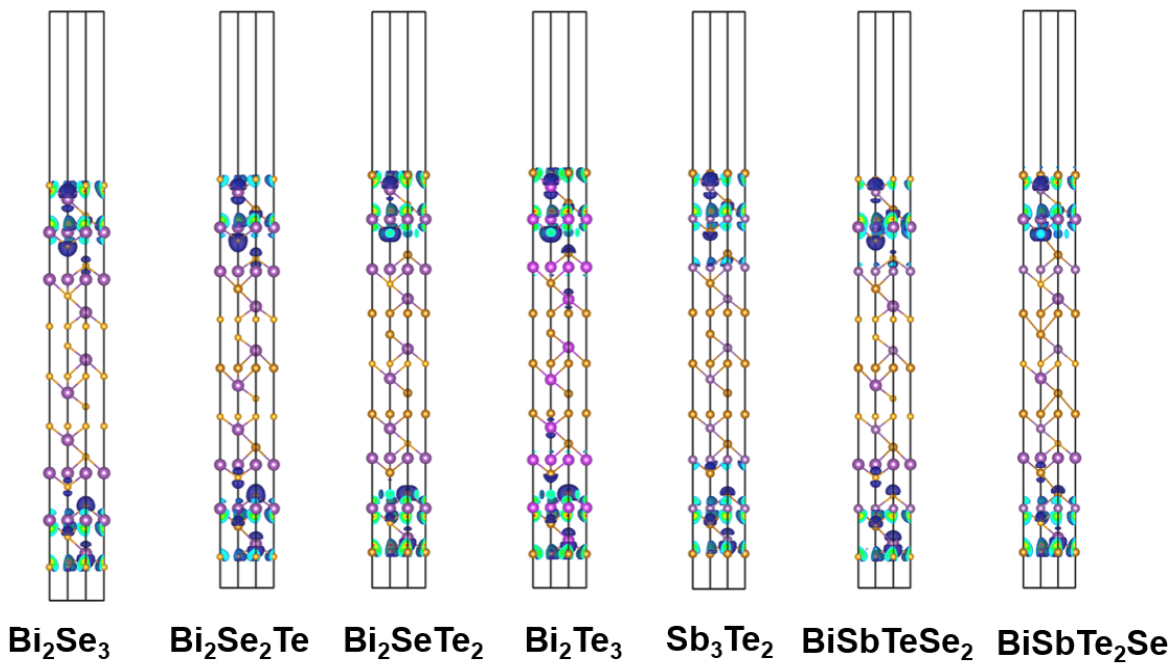

**Supplementary Figure 9.** Three-dimensional charge density distribution of conducting surface states for BSTS compounds with isovalue of 0.005 e/Å$^3$.

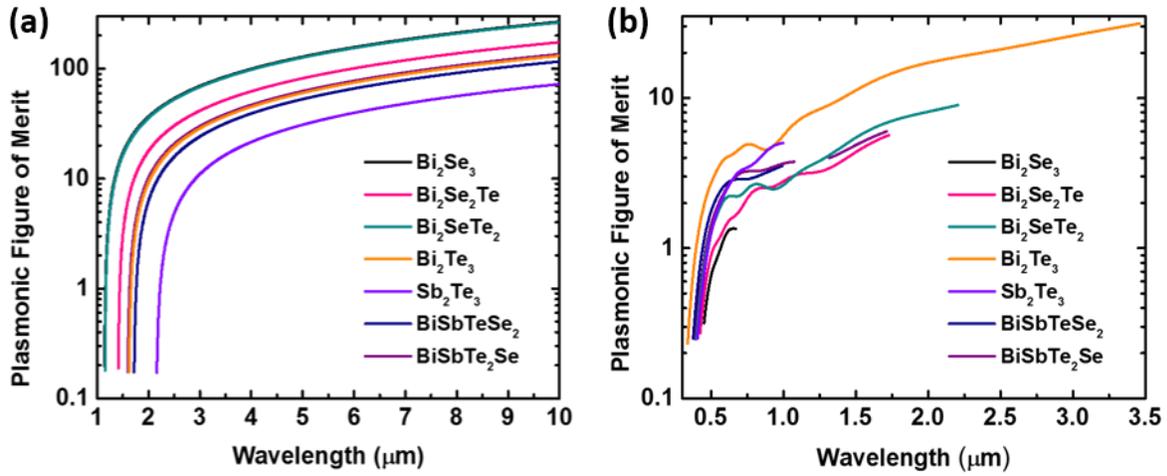

**Supplementary Figure 10.** Figures of merit of plasmonic response of various BSTS topological insulator slabs arising due to (a) 2D Drude response of surface free carriers (intraband contribution), and (b) interband transitions between surface states and higher energy states.